\documentclass[aps,prl,nofootinbib,twocolumn,superscriptaddress,preprintnumbers]{revtex4-2} 
\usepackage[nodisplayskipstretch]{setspace}
\usepackage{cancel}
\usepackage{mathtools}

\usepackage{amssymb}
\usepackage{amsmath}
\usepackage{epsfig}
\usepackage{hyperref}
\usepackage{breakurl}
\usepackage{bm}
\usepackage{color}
\usepackage{slashed}
\usepackage{tikz}
\usetikzlibrary{decorations.markings}
\usepackage[english]{babel}
\newcommand\beq{\begin{equation}}
\newcommand\eeq{\end{equation}}
\def\bea{\begin{eqnarray}}
\def\eea{\end{eqnarray}}

\DeclareRobustCommand{\SkipTocEntry}[4]{}

\newcommand{\nn}{\nonumber}

\newcommand\beal{\begin{aligned}}
\newcommand\eeal{\end{aligned}}

\usepackage{xcolor}

\newcommand{\bp}{{\boldsymbol p}}

\newcommand\cE{\mathcal{E}}


\begin{document}

\preprint{DESY\, 21-226\\\phantom{~}}
\title{Conservative Dynamics of Binary Systems at Fourth\\ [0.2cm]  Post-Minkowskian Order in the Large-eccentricity Expansion }
\author{Christoph Dlapa}
\author{Gregor K\"alin}
\author{Zhengwen Liu}
\author{Rafael A. Porto}
\affiliation{ Deutsches Elektronen-Synchrotron DESY, Notkestr. 85, 22607 Hamburg, Germany}

\begin{abstract}
We compute the conservative dynamics of non-spinning binaries at fourth Post-Minkowskian order in the large-eccentricity limit, including both potential and radiation-reaction tail effects. This is achieved by obtaining the scattering angle in the worldline effective field theory  approach and deriving the bound radial action via analytic continuation. The~associated integrals are bootstrapped to all orders in velocities through differential equations, with boundary conditions in the potential and radiation regions. The large angular momentum expansion captures all the local-in-time effects as well as the trademark logarithmic corrections for generic bound orbits. Agreement is found in the overlap with the state-of-the-art in Post-Newtonian theory.  \end{abstract}
 
\maketitle

\emph{Introduction.} The era of gravitational wave (GW) science began in spectacular fashion with several detections already reported by the LIGO-Virgo-KAGRA collaboration \cite{LIGOScientific:2021djp}, and many more yet to come with the future planned observatories such as LISA \cite{LISA} and the Einstein Telescope \cite{ET}. Motivated by the initial breakthroughs and the expected scientific output \cite{buosathya,tune,music,Maggiore:2019uih,Barausse:2020rsu}, a community effort has been established toward constructing high-accurate waveform models for the emission of GWs from binary systems. This includes numerical simulations for the merger phase \cite{Ajith:2012az,Szilagyi:2015rwa,Dietrich:2018phi} as well as analytic techniques to tackle the inspiral in the Post-Newtonian (PN) regime, using both traditional \cite{blanchet,Schafer:2018kuf} and modern methodologies such as the effective field theory (EFT) approach~\cite{walterLH,iragrg,foffa,review}.\vskip 4pt These developments, in particular the use of tools from particle physics pioneered in \cite{nrgr}, have impacted our understanding of the two-body problem in PN theory,  e.g. \cite{nrgrs,dis1,dis2,prl,nrgrss,nrgrs2,andirad,andirad2,amps,srad,chadRR,natalia1,natalia2,tail,apparent,lamb,nrgr4pn1,nrgr4pn2,tail3,5pn1,5pn2,hered1,hered2,Blumlein:2020pyo,Blumlein:2021txe,foffa3,blum,blum2,dis3,Levi:2020uwu,Levi:2020kvb,withchad,radnrgr,pardo,Cho:2021mqw,Cho:2022syn}, leading to the knowledge of the conservative spin-independent gravitational potential at fourth PN (4PN) order \cite{tail,apparent,lamb,nrgr4pn1,nrgr4pn2}, in parallel with independent derivations using traditional methods \cite{Damour:2014jta,Jaranowski:2015lha,Bernard:2015njp, Bernard:2017bvn,Marchand:2017pir,damour3n,binidam1,binidam2,Bini:2021gat}. The current state-of-the-art includes reports of contributions at 5PN \cite{5pn1,5pn2,Blumlein:2020pyo,Blumlein:2021txe,hered1,hered2}, and partial results at 6PN  \cite{blum,blum2,binidam1,binidam2,Bini:2021gat}.\vskip 4pt

Inspired by the EFT framework in the PN regime~\cite{walterLH,iragrg,foffa,review}, novel ideas from the theory of scattering amplitudes~\cite{reviewdc}, and the existence of a correspondence for observables in hyperbolic-like and elliptic-like orbits via analytic continuation in the binding energy and angular momentum---dubbed the {\it Boundary-to-Bound} (B2B) correspondence \cite{paper1,paper2}---significant efforts have been invested in recent years towards studying scattering processes within the Post-Minkowskian (PM) expansion, both with amplitude-based~\cite{ira1,Vaidya:2014kza,cheung,Guevara:2018wpp,donal,donalvines,cristof1,bohr,zvi1,zvi2,paper1,paper2,Haddad:2020que, Aoude:2020onz,parra,zvispin,soloncheung,andres2,4pmzvi,Kreer:2021sdt,Parra2,Gabriele,DiVecchia:2021bdo,Cristofoli:2021vyo,Bautista:2021wfy} and EFT-based \cite{Goldberger:2016iau,Walter,pmeft,3pmeft,tidaleft,pmefts,janmogul,janmogul2,Mougiakakos:2021ckm,4pmeft,Jakobsen:2021lvp,Riva:2021vnj} methodologies. The PM expansion naturally encapsulates an infinite (resummed) series of velocity corrections at each order in Newton's constant, which may result in improved waveform models,~e.g.~\cite{Antonelli:2019ytb}.\vskip 4pt After the seminal result at third PM (3PM) order for non-spinning binary systems \cite{zvi1,zvi2,3pmeft}, partial (potential-only) corrections at 4PM have been obtained within both approaches \cite{4pmzvi,4pmeft}. Here we extend the knowledge of the two-body dynamics at ${\cal O}(G^4)$, by including both potential and radiation-reaction tail effects---the latter being due to the back-scattering of the outgoing GWs on the background geometry---thus removing spurious divergent terms in previous potential-only computations.  Similarly to the Lamb~shift~\cite{lamb}, yet in a classical setting, mode-factorization into potential (off-shell) and radiation (on-shell) regions led to infrared(IR)/ultraviolet(UV) divergences in PN computations \cite{Damour:2014jta,Jaranowski:2015lha,Bernard:2015njp, Bernard:2017bvn,Marchand:2017pir}, which ultimately cancel out in physical observables~\cite{tail,apparent,nrgr4pn2}. As we demonstrate here, the explicit cancelation is also manifest in the PM regime, yielding (ambiguity-free) finite results.\vskip 4pt 

Our derivation proceeds through the scattering angle computed in the EFT approach \cite{pmeft,3pmeft}, in conjunction with the B2B map \cite{paper1,paper2} between unbound and bound motion extended to radiative effects~\cite{b2b3}.  Using multi-loop integration tools from particle physics \cite{Smirnov:2012gma,Kotikov:1991pm,Remiddi:1997ny,Henn:2013pwa,Prausa:2017ltv,Lee:2020zfb,Lee:2014ioa,Adams:2018yfj,Chetyrkin:1981qh,Tkachov:1981wb,Smirnov:2019qkx,Smirnov:2020quc,Lee:2012cn,Lee:2013mka,Beneke:1997zp,Jantzen:2012mw,Smirnov:2015mct,Meyer:2016zeb,Meyer:2016slj,Broedel:2019kmn,Primo:2017ipr,Hidding:2020ytt,Goncharov:2001iea,Chen:1977oja,Duhr:2014woa,Duhr:2019tlz,Dlapa:2020cwj,Smirnov:2021rhf,Lee:2019zop},  the calculation is reduced to a series of `three-loop'  (massless) integrals which are computed through differential equations. The resulting deflection angle features {\it logarithm}, {\it dilogarithm} and {\it complete elliptic integrals} of the first and second kind, and agrees in the overlap with the state-of-the-art in PN theory 
\cite{binidam1,binidam2,hered1,hered2,Blumlein:2020pyo,Blumlein:2021txe}. For~completeness, we reconstruct the center-of-mass momentum.  

\vskip 4pt

\begin{figure}[t!] 
\includegraphics[width=0.32\textwidth]{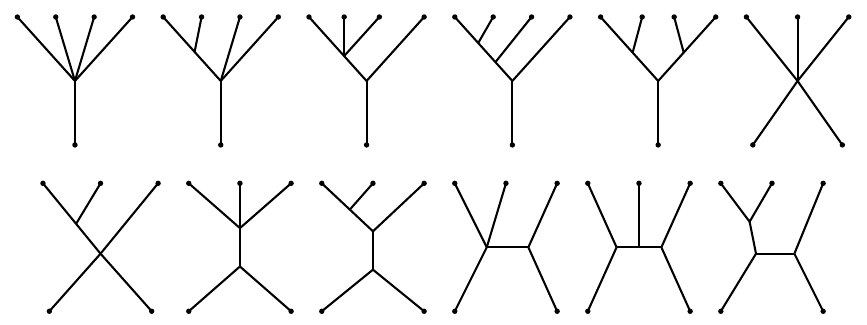} 
\includegraphics[width=0.32\textwidth]{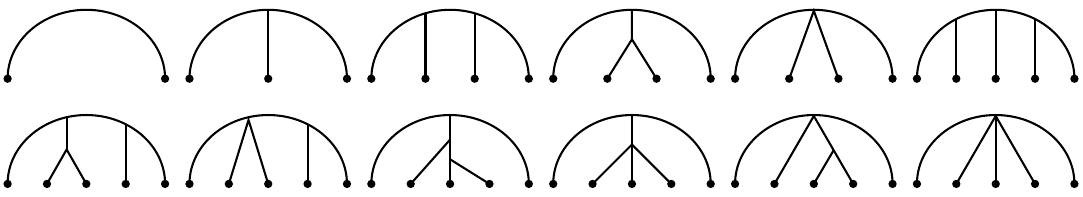} 
      \caption{Feynman topologies needed to compute the deflection angle at ${\cal O}(G^4)$. The solid lines represent the gravitational field and the dots account for the worldline sources. The ones in the third and fourth row are {\it self-energy} diagrams (involving only a single particle) needed when radiation fields are included.} 
      \label{topo}
            \vspace{-0.4cm}
\end{figure}

 {\it The EFT formalism.} Following the procedure discussed in \cite{pmeft,3pmeft,4pmeft}, the effective action $(S_{\rm eff})$ is obtained by \emph{integrating out} the metric perturbation, $h_{\mu\nu} = g_{\mu\nu}-\eta_{\mu\nu}$, using the (classical) saddle-point approximation of the path integral, schematically \beq \int Dh \, e^{i\left (S_{\rm EH}+S_{\rm pp}\right )} \to e^{iS_{\rm eff}}\,,\eeq  in Einstein's gravity ($S_{\rm EH}$) coupled to point-like worldline sources $(S_{\rm pp})$, ignoring quantum effects. The computation is reduced to a series of (`tree-level') Feynman diagrams. The~full set of topologies at 4PM are shown in Fig.~\ref{topo}.  As before \cite{4pmeft}, we~must include mirror images and {\it iterations} from lower order solutions to the trajectories.\vskip 4pt 
We restrict ourselves here to potential modes and conservative tail terms. The~latter entail Feynman diagrams with only two radiation modes coupled to a background potential.  In~this scenario, the conservative contribution is captured by the standard Feynman rules and $i0$-prescription for the propagators of the gravitational field, i.e. $\tfrac{i}{p_0^2-\bp^2+i0}$, as long as we consider the real part of the effective action~\cite{tail,foffa3}, ignoring dissipative effects. This allows us to retain the integration machinery intact. \vskip 4pt
From the resulting effective action we can then compute the scattering angle, $\chi$, through the impulse, $\Delta p_{a=1,2}^\mu$, evaluated on the classical trajectory \cite{pmeft,3pmeft,4pmeft}, \beq
2\sin(\chi/2) = \sqrt{-\Delta p_a^2}/p_\infty\,,\label{angle}
\eeq 
with $p_\infty$ the incoming momentum in the center-of-mass frame.  As we mentioned, spurious IR/UV divergences appear due to mode factorization, e.g.~\cite{apparent}. Hence, we work in dimensional regularization and write the intermediate result for the PM expansion of the angle in $d=4-2\epsilon$ dimensions as \cite{4pmeft}
\beq
\frac{\chi}{2} =  \sum_n \left(\left(4\bar\mu^2b^2\right)^\epsilon \frac{GM}{b}\right)^n \chi^{(n)}_{b}\,,\label{pmexp}
\eeq
where $b = \sqrt{-b^\mu b_\mu}$ is the impact parameter, $\chi^{(n)}_b$ the associated PM coefficients, $\bar\mu^2 \equiv 4\pi \mu^2 e^{\gamma_E}$ the renormalization scale (with $\gamma_E$ Euler's constant), and $M=m_1+m_2$ the total mass. As expected, the combined result is devoid of divergences or ambiguities \cite{apparent,lamb}.\vskip 4pt

{\it Integration.} The scattering angle can be reduced to the computation of the following set of  (three-loop) integrals,
\begin{align}\label{4pm-ints}
\prod_{i=1}^3 \int \frac{\mathrm{d}^d \ell_i}{\pi^{d/2}}
\frac{\delta(\ell_i \!\cdot\! u_{a_i})}  { 
(\pm \ell_i \!\cdot\! u_{\slashed{a}_i} {-} i0)^{n_i}} {1 \over \prod_{j=1}^{9} D_j^{\nu_j}}\,,
\end{align}
restricted by Dirac-$\delta$ functions, where $\ell_{i=1,2,3}$ are the loop momenta, $n_i,\nu_j$ are integers, $a_i\in\{1,2\}$ ($u_\slashed{1}=u_2$, $u_\slashed{2}=u_1$), and $D_j$ are various sets of quadratic propagators: $\{\ell_i^2+i0,(\ell_i-q)^2+i0,\ldots\}$. The~external data~obeys $q\cdot u_a=0$ and $u_a^2=1$, with $q$ the transfer momentum and $u_{a=1,2}$ the incoming velocities. Hence, after factoring out the dependence on $q^2$ using dimensional analysis, the result of the integrals can only be a function of~$\gamma\equiv u_1\cdot u_2$.\vskip 4pt Following our previous derivations \cite{3pmeft,4pmeft}, introducing the parameter $x$ defined through $\gamma \equiv (x^2 {+} 1)/2x$~\cite{parra}, the value of these integrals is obtained by the method of differential equations \cite{Kotikov:1991pm,Remiddi:1997ny,Henn:2013pwa,Lee:2014ioa, Prausa:2017ltv, Adams:2018yfj, Lee:2020zfb}, with boundary conditions computed in the near-static limit $x\simeq 1$. We make extensive use of the integration-by-parts (IBP) relations \cite{Chetyrkin:1981qh, Tkachov:1981wb}, via \texttt{FIRE6} \cite{Smirnov:2019qkx,Smirnov:2020quc} and \texttt{LiteRed} \cite{Lee:2012cn,Lee:2013mka}, as well as the   \texttt{INITIAL} algorithm \cite{Dlapa:2020cwj}.  The final result for the deflection angle involves logarithms, dilogarithms ($\text{Li}_2(x)$), as well as complete elliptic integrals of the first $(\mathrm{K}(x))$ and second ($\mathrm{E}(x)$) kind \cite{4pmzvi,4pmeft}. (See \cite{dklp} for more details in the integration procedure.) 
\vskip 4pt  

{\it Potential Region.} As a check, we re-evaluated the boundary conditions for the differential equations  at $x=1$ with potential modes. As discussed in \cite{4pmeft}, these may be reduced into a basis containing only seven integrals via additional IBP identities, which we computed by direct integration. As~expected, the self-energy diagrams in Fig.~\ref{topo} turn into (scaleless) integrals which vanish in dimensional regularization, and therefore do not contribute in the potential region. Adding the pieces together we recover the result~in~\cite{4pmzvi,4pmeft} 
\beq
\begin{aligned}
{\chi^{(4)}_{b\, (\rm pot)} \over \pi \Gamma}=  \chi_s(x) + \nu\left( \frac{\chi_{2\epsilon}(x)}{2\epsilon} + \chi_p(x)\right) \,,\label{potangle}
\end{aligned}
\eeq
for the potential contribution to the scattering angle at ${\cal O}(G^4)$, where  $\Gamma \equiv E/M$, with $E$ the total energy, and $\nu=m_1m_2/M^2$ the symmetric mass-ratio. Expressions for the $(\chi_s,\chi_{2\epsilon},\chi_p)$ coefficients are given in~\cite{4pmeft} and the ancillary file, see also \eqref{totangle2} and the supplemental material.
\vskip 4pt

{\it Tail Region.}  The boundary conditions including radiation-reaction effects is more challenging, mainly due to the interplay between potential and radiation modes. We use the {\it asy2.m} code in the \texttt{FIESTA} package to identify the relevant regions of integration \cite{Jantzen:2012mw,Smirnov:2015mct,Smirnov:2021rhf}. We~find several contributions featuring one, two and up to three radiation modes. We~keep only regions consistent with conservative radiation-reaction tail effects.\vskip 4pt  
 After performing a Laurent expansion around $x\simeq 1$, yielding the anticipated pole in $(1-x)^{-4\epsilon}$~\cite{4pmeft}, we computed the associated boundary integrals using various consistency relations \cite{Lee:2019zop}. Collecting the terms we~find
\beq
\begin{aligned}
{\chi^{(4)}_{b\, (\rm tail)} \over \pi \Gamma}=  \nu\left( -\frac{\chi_{2\epsilon}(x)}{2\epsilon} (1-x)^{-4\epsilon} + \chi_t(x)\right) \,,\label{radangle}
\end{aligned}
\eeq 
for the (conservative) contribution to the deflection angle due to tail effects at 4PM. The value of $\chi_t(x)$ is given in the supplemental material and ancillary file. 
\begin{widetext}

{\it Combined Result.} As expected, the divergence and $\mu$-dependence cancel out and the combined result becomes
\beq
\begin{aligned}
\frac{\chi^{(4)}_{b\, (\rm comb)}}{\pi \Gamma} =   \chi_s + \nu \Big(\chi_c (x)+2\chi_{2\epsilon}(x)\log (1-x)\, \Big) \,,\label{totangle}
\end{aligned}
\eeq
where
\begin{equation}
\allowdisplaybreaks
\label{totangle2}
  \small
  \begin{aligned}
     \chi_s(x) &= \frac{105 h_1(x)}{128 \left(x^2-1\right)^4}\,,\quad
    \chi_{2\epsilon}(x) = 
    -\frac{3 h_2(x) \log (x)}{32 x \left(x^2-1\right)^5}
    +\frac{3 h_3(x) \log \left(\frac{x+1}{2}\right)}{32 x^2 \left(x^2-1\right)^2}
    +\frac{h_4(x)}{64 x^2 \left(x^2-1\right)^4}\,,\\
  \chi_c(x) &=
    -\frac{21 h_6(x) \mathrm{E}^2\left(1-x^2\right)}{8 \left(x^2-1\right)^4}
    +\frac{3 h_7(x) \mathrm{K}\left(1-x^2\right) \mathrm{E}\left(1-x^2\right)}{8 \left(x^2-1\right)^4}
    -\frac{15 h_8(x) \mathrm{K}^2\left(1-x^2\right)}{16 \left(x^2-1\right)^4}
    -\frac{h_{16}(x) \log \left(x^2+1\right)}{32 x^3 \left(x^2-1\right)^4}\\
    &\quad+\frac{3 h_{19}(x) \text{Li}_2\left(-\frac{(x-1)^2}{(x+1)^2}\right)}{128 x^4 \left(x^2-1\right)^2}
    +\frac{\pi ^2 h_{35}(x)}{512 (x-1)^3 x^4 (x+1)^5}
    +\frac{3 h_{36}(x) \log ^2(2) }{16 x^2 \left(x^2-1\right)^2}
    +\frac{3h_{37}(x) \log (2) \log (x)}{8 \left(x^2-1\right)^5}
    -\frac{3 h_{38}(x) \log (2) \log (x+1)}{16 x^2 \left(x^2-1\right)^2}\\
    &\quad+\frac{3 h_{39}(x) \log (2) }{16 x^2 \left(x^2-1\right)^4}
    +\frac{3 h_{40}(x) \log ^2(x)}{256 x^4 \left(x^2-1\right)^8}
    -\frac{3 h_{41}(x) \log (x) \log (x+1)}{128 x^4 \left(x^2-1\right)^5}
    +\frac{h_{42}(x) \log (x)}{64 x^3 \left(x^2-1\right)^7}
    -\frac{3 h_{43}(x) \log ^2(x+1)}{2 x \left(x^2-1\right)^2}\\
    &\quad+\frac{h_{44}(x) \log (x+1)}{32 x^3 \left(x^2-1\right)^4}
    +\frac{3 h_{45}(x) \left(\text{Li}_2\left(\frac{x-1}{x}\right)-\text{Li}_2(-x)\right)}{128 (x-1)^3 x^4 (x+1)^5}
    -\frac{3 h_{46}(x) \text{Li}_2\left(\frac{x-1}{x+1}\right)}{64 (x-1)^2 x^4}
    +\frac{h_{47}(x)}{384 x^3 \left(x^2-1\right)^6 \left(x^2+1\right)^7}
   \,, \end{aligned}
   \end{equation}
\end{widetext} 
with the explicit value of the $h_i(x)$ polynomials displayed in the ancillary file (see also the supplemental material for the intermediate results leading to \eqref{totangle2}).  After expanding in small velocities we find perfect agreement with the PN state-of-the-art value reported in \cite{Blumlein:2020pyo,Blumlein:2021txe,hered1,hered2,binidam1,binidam2}.\footnote{There is a mismatch at ${\cal O}(\nu^2)$ between the results in \cite{Blumlein:2020pyo,Blumlein:2021txe,hered1,hered2} and those in \cite{binidam1,binidam2}, which can be traced to the definition of conservative terms in \cite{binidam1,binidam2}.} 

\vskip 4pt    {\it Boundary-to-Bound correspondence.} As it was shown in \cite{paper1,paper2,b2b3}, the B2B dictionary allows us to derive PM-expanded observables for bound orbits from the scattering angle computed in a large-eccentricity (or large angular momentum) expansion. After analytic continuation in angular momentum, and to negative binding energies, we obtain \beq
 i^{\rm 4PM}_{r}  =   \frac{2 (1-\gamma^2)^2 }{3(\Gamma j)^3} \left[\chi_s  + \nu \left(\chi_c +\chi_{2\epsilon}\log (x-1)^2\, \right) \right] 
 \label{ptir}
\eeq
for the B2B large-eccentricity approximation to the (reduced) 4PM bound radial action, with $j \equiv J/(GM^2\nu)$ the (dimensionless) angular momentum. From the radial action we can then derive all the observables for elliptic-like orbits through differentiation, including the binding energy which is one of the main ingredients needed to compute the GW phase evolution in an adiabatic approximation~\cite{blanchet,Antonelli:2019ytb}, providing an infinite series of velocity corrections at ${\cal O}(G^4)$.\vskip 4pt As it is known from the PN literature, e.g.~\cite{tail,binidam1,binidam2}, tail terms feature both local- as well as non-local-in-time dynamical effects. As it was discussed in \cite{paper1,paper2,b2b3}, the expression in \eqref{ptir} readily captures all local-in-time contributions to gauge-invariant observables for generic bound orbits (also for aligned-spin effects). Remarkably, the same is true for the trademark (non-local) logarithmic tail corrections, which may be obtained via \cite{4pmeft}
\beq
\begin{aligned}
i^{\rm 4PM}_{r (\rm log)} &=   \frac{2\nu}{3} \frac{(1-\gamma^2)^2}{(\Gamma j)^3}   \chi_{2\epsilon}(\gamma)  \log |{\cal E}|\,,
\end{aligned}
\eeq
to all orders in velocity, with ${\cal E}\equiv \tfrac{E-M}{M\nu}$ the (reduced) binding energy. This is not the case, however, with other non-local-in-time effects for generic orbits, which do not transition smoothly between hyperbolic- and elliptic-like motion and therefore cannot be derived from the knowledge of the scattering angle \cite{b2b3}.\vskip 4pt From~the deflection angle we can also reconstruct the 4PM coefficient, $f_4(\cE)$, of the PM expansion of an effective (local-in-time) center-of-mass momentum
\beq
\bp^2 = p_\infty^2 \left[1+ \sum_{n=1}^\infty f_n(\cE) \left(\frac{GM}{r}\right)^n\right]\,,\label{pcom}
\eeq
 in an isotropic gauge, such that $i_r \propto \int p_r dr$, using the relationship \cite{paper1,paper2}
\beq
\begin{aligned}
f_4 = &\frac{8}{3\pi} \chi^{(4)}_{b} - 2 \chi^{(3)}_{b} \chi^{(1)}_{b} -\frac{8}{\pi^2} (\chi^{(2)}_b)^2 \\
&+\frac{8}{\pi} (\chi^{(1)}_{b})^2 \chi^{(2)}_{b}-\frac{2}{3}  (\chi^{(1)}_{b})^4\,,
\end{aligned}
\eeq
together with previous results to 3PM \cite{pmeft,3pmeft}. Explicit values are given in the ancillary file. The Hamiltonian can be also reconstructed using the algebraic relations in \cite{paper1,paper2}. Notice, as advertised in \cite{4pmeft,b2b3}, that in all cases the factors of $\log r$ in the intermediate results drop out of the final expressions.\vskip 8pt   {\it Conclusions.} We have computed the contribution from potential and radiation-reaction tail effects to the conservative dynamics of binary compact objects to 4PM order in the large-eccentricity limit. Our (ambiguity-free) result for the deflection angle at 4PM agrees in the overlap with state-of-the-art computations in PN theory for the combined potential and tail contributions \cite{Blumlein:2020pyo,Blumlein:2021txe,hered1,hered2,binidam1,binidam2}. As~it was already the case in previous derivations in the EFT approach \cite{3pmeft,4pmeft}, the PM result can be entirely bootstrapped from PN data to all orders in velocities through differential equations---at this order including a sector involving elliptic integrals---without resorting to PN resummations. The boundary conditions (in the near-static limit) were obtained via the method of regions with potential and radiation modes.\vskip 4pt  There are, however, some important caveats that need to be addressed in order to complete the knowledge of the conservative 4PM dynamics for generic orbits, notably the mapping between unbound and bound motion for all the non-local-in-time effects. 
Moreover, there is also the issue of conservative non-linear memory terms.\footnote{The result in \cite{Blumlein:2021txe}  suggests the appearance of a conservative memory term in the overlap between 4PM and 5PN orders~at~${\cal O}(\nu^2)$.} The~latter arise from the interaction between the outgoing GW radiation and the waves emitted by the binary system at an earlier time. The derivation of memory effects at 4PM,  the extension of the B2B map to generic non-local-in-time terms, as well as the computation of higher PM orders, is underway in the EFT approach. \vskip 8pt

 {\it Acknowledgements.} We thank Johannes Bl\"umlein, Ekta Chaubey, Gihyuk Cho, Stefano Foffa, Ryusuke Jinno, Francois Larrouturou, Andreas Maier, Massimiliano Riva, Henrique Rubira, Riccardo Sturani, Yang Zhang, and Simone Zoia for useful conversations. This work received support from the ERC-CoG  {\it Precision Gravity: From the LHC to LISA} provided by the European Research Council (ERC) under the European Union's H2020 research and innovation programme (grant No.\,817791). The research of C.D. was  funded in part by the ERC-CoG grant  {\it Novel structures in scattering amplitudes} (grant No.\,725110).

\appendix
\begin{widetext}

\section{Supplemental Material}

The coefficients of the scattering angle for the potential (beyond the test-body limit) and tail contributions to 4PM order displayed in the text are given by:\vskip 12pt
\begin{equation}
  \small
  \begin{aligned}
    \chi_p(x) &= 
    \frac{\pi ^2 h_5(x)}{1024 x^4 \left(x^2-1\right)^5}
    -\frac{21 h_6(x) \mathrm{E}^2\left(1-x^2\right)}{8 \left(x^2-1\right)^4}
    +\frac{3 h_7(x) \mathrm{K}\left(1-x^2\right)  \mathrm{E}\left(1-x^2\right)}{8 \left(x^2-1\right)^4}
    -\frac{15 h_8(x) \mathrm{K}^2\left(1-x^2\right)}{16 \left(x^2-1\right)^4}
    +\frac{3 h_9(x) \log^2\left(\frac{x+1}{2}\right)}{16 x^2 \left(x^2-1\right)^2}\\
    &\quad+\frac{3h_{10}(x) \log (2) \log (x)}{8 \left(x^2-1\right)^5}
    +\frac{h_{11}(x) \log (2)}{128 x^3 \left(x^2-1\right)^4}
    +\frac{3 h_{12}(x) \log ^2(x)}{512 x^4 \left(x^2-1\right)^8}
    -\frac{3 h_{13}(x) \log (x) \log (x+1)}{256 x^4 \left(x^2-1\right)^5}
    +\frac{h_{14}(x) \log (x)}{128 x^3 \left(x^2-1\right)^7}\\
    &\quad-\frac{h_{15}(x) \log (x+1)}{128 x^3 \left(x^2-1\right)^4}
    -\frac{h_{16}(x) \log \left(x^2+1\right)}{64 x^3 \left(x^2-1\right)^4}
    +\frac{3 h_{17}(x) \text{Li}_2\left(\frac{x-1}{x}\right)}{256 x^4 \left(x^2-1\right)^5}
    -\frac{3 h_{18}(x) \text{Li}_2(-x)}{256 x^4 \left(x^2-1\right)^5}
    +\frac{3 h_{19}(x) \text{Li}_2\left(-\frac{(x-1)^2}{(x+1)^2}\right)}{256 x^4 \left(x^2-1\right)^2}\\
    &\quad-\frac{3 h_{20}(x)\text{Li}_2\left(\frac{x-1}{x+1}\right)}{128 x^4 \left(x^2-1\right)^2}
    +\frac{h_{21}(x)}{1536 x^3 \left(x^2-1\right)^6 \left(x^2+1\right)^7}
    \,,\\
    \chi_t(x) &=
    -\frac{h_{16}(x) \log \left(x^2+1\right)}{64 x^3 \left(x^2-1\right)^4}
    +\frac{3 h_{19}(x) \text{Li}_2\left(-\frac{(x-1)^2}{(x+1)^2}\right)}{256 x^4 \left(x^2-1\right)^2}
    +\frac{h_{22}(x) \left(12 \text{Li}_2(-x)+\pi ^2\right)}{1024 x^4 \left(x^2-1\right)^5}
    -\frac{24 h_{23}(x) \log ^2(2) }{\left(x^2-1\right)^2}
    -\frac{6 h_{24}(x) \log (2) \log (x)}{\left(x^2-1\right)^5}\\
    &\quad+\frac{3 h_{25}(x) \log (2) \log (x+1)}{16 x^2 \left(x^2-1\right)^2}
    -\frac{h_{26}(x) \log (2)}{128 x^3 \left(x^2-1\right)^4}
    -\frac{3 h_{27}(x) \log ^2(x)}{512 x^4 \left(x^2-1\right)^8}
    +\frac{3 h_{28}(x) \log (x+1) \log (x)}{256 x^4 \left(x^2-1\right)^5}
    -\frac{h_{29}(x) \log (x)}{128 x^3 \left(x^2-1\right)^7}\\
    &\quad-\frac{3 h_{30}(x) \log ^2(x+1)}{16 x^2 \left(x^2-1\right)^2}
    +\frac{h_{31}(x) \log (x+1)}{128 x^3 \left(x^2-1\right)^4}
    -\frac{3 h_{32}(x) \text{Li}_2\left(\frac{x-1}{x}\right)}{256 x^4 \left(x^2-1\right)^5}
    +\frac{3 h_{33}(x) \text{Li}_2\left(\frac{x-1}{x+1}\right)}{128 x^4 \left(x^2-1\right)^2}
    +\frac{h_{34}(x)}{1536 x^3 \left(x^2-1\right)^6 \left(x^2+1\right)^7}\nn
    \,,  \end{aligned}
\end{equation}
\vskip 12pt
\noindent where the $h_i(x)$'s are polynomials in $x$ up to degree 32, collected in the ancillary file. We use the following~conventions 
\beq
\begin{aligned}
\text{Li}_2(z) &\equiv \int_z^0 dt \, \frac{\log(1-t)}{t}\,,   \,\\
\mathrm{K}(z) &\equiv 
 \int_0^1 \frac{dt}{\sqrt{\left(1-t^2\right)\left(1-z t^2\right)}}\,,\\
\mathrm{E}(z) &\equiv  \int_0^1 dt\, \frac{\sqrt{1-z t^2}}{\sqrt{1-t^2}}\,,\nn
\end{aligned}
\eeq
for the dilogarithm, and complete elliptic integral of the first and second kind, respectively.
\end{widetext}
\bibliography{ref4PM}
\end{document}